\documentclass[12pt]{iopart}

\usepackage{graphicx}
\usepackage{mathrsfs}
\usepackage{bbm}

\def\egp{\ensuremath{e_g^\pi}}
\def\egs{\ensuremath{e_g^\sigma}}
\def\t2g{\ensuremath{t_{2g}}}
\def\a1g{\ensuremath{a_{1g}}}

\newcommand{\eg}{%
          {e.g.\ }}
\newcommand{\ie}{%
          {i.e.\ }}

\newcommand{\svek}{%
        \mathbf}
\newcommand{\vek}[1]{%
        \hbox{\textbf #1}}

\newcommand{\op}[1]{%
        \hbox{\textbf #1}}
\newcommand{\pr}{%
        ^\prime}

\newcommand{\dif}{%
        \hbox{d}}  

\newcommand{\bra}[1]{\ensuremath{\langle #1|}}
\newcommand{\ket}[1]{\ensuremath{|#1\rangle}}
\newcommand{\braket}[2]{\langle #1|#2\rangle}

\newcommand{\im}{%
        {\imath}}    
        
\usepackage{iopams}  

\begin{document}

\title{Multi-orbital Effects in Optical Properties of Vanadium Sesquioxide}

\author{Jan M. Tomczak}
\address{Research Institute for Computational Sciences, AIST, Tsukuba, 305-8568 Japan}
\address{Japan Science and Technology Agency, CREST}

\author{Silke Biermann}
\address{Centre de Physique Th{\'e}orique, Ecole Polytechnique,
91128 Palaiseau Cedex, France}
\address{Japan Science and Technology Agency, CREST}

\ead{jan.tomczak@polytechnique.edu}

\begin{abstract}
Vanadium sesquioxide, V$_2$O$_3$, boasts a rich phase diagram whose description necessitates the accounting for many-body Coulomb correlations.
Spectral properties of this compound have been successfully addressed within
dynamical mean field theory to an extent that results of recent angle resolved photoemission experiments have been correctly predicted. 
While photoemission spectroscopy probes the occupied part of the one-particle spectrum, optical experiments measure transitions into empty
states and thus provide complementary information. In this work, we focus on the optical properties of V$_2$O$_3$ in its
paramagnetic phases by employing our recently developed ``generalized Peierls approach''. 
Compared to experiments, we obtain results in overall satisfactory 
agreement. Further, we rationalize that the experimentally observed temperature dependence stems from the different coherence scales of the involved charge carriers.
\end{abstract}

%Uncomment for PACS numbers title message
%\pacs{00.00, 20.00, 42.10}
% Keywords required only for MST, PB, PMB, PM, JOA, JOB? 
%\vspace{2pc}
%\noindent{\it Keywords}: Article preparation, IOP journals
% Uncomment for Submitted to journal title message
%\submitto{\JPB}

% Comment out if separate title page not required
%\maketitle

\section{Introduction}

Vanadium sesquioxide, V$_2$O$_3$, has been the subject of extensive theoretical and
experimental studies for now more than three decades.
It is considered as {\it the} prototype compound, that undergoes
a Mott-Hubbard transition~\cite{mott,imada} in its purest form. Indeed, the high-temperature
(T$>$T$_{\hbox{\small N\'eel}}$) metal-insulator transition upon chemical substitution,
$($V$_{1-x}$Cr$_x)_2$O$_3$, is isostructural and no magnetic order is acquired.
Early theoretical approaches  resorted to the
Hubbard model to explain the  electronic properties of V$_2$O$_3$.
However, over the years, experiments indicated  that the physics
 of this material is more involved and a
realistic multi-orbital setup is needed for the 
complexity of the correlation effects taking place.
For reviews see \eg~\cite{mott,imada,me_phd}.

The field of correlated materials gained major momentum from the development of 
dynamical mean field theory (DMFT)~\cite{bible}.
In combination with standard density functional based methods like the local density approximation (LDA)
the calculation of spectral properties of materials with
strong electronic Coulomb interactions became possible.
Over the past years, LDA+DMFT enlightened our understanding of materials such as
transition metals, their oxides or sulphides, as
well as f-electron compounds~\cite{vollkot}.
Several works highlighted the applicability of the technique to V$_2$O$_3$~\cite{PhysRevLett.86.5345,keller:205116, PhysRevLett.91.156402,laad:045109,anisimov:125119,poter_v2o3}.
In our previous work~\cite{poter_v2o3}, we find 
that the metal-insulator transition is not due to the 
Brinkman-Rice mechanism~\cite{PhysRevB.2.4302} in its single-band form, but results
from the 
impact of Coulomb correlations on the crystal-field splitting.
Owing to its octahedral oxygen surrounding, the vanadium 3d-orbitals
split into two \egs\ and three lower
lying \t2g\ orbitals.
The two manifolds of bands
are isolated in energy both from each other and from other orbitals. The trigonal part of the crystal field further
splits the \t2g\ into an \a1g\  and two lower lying degenerate \egp\ orbitals.
The local Coloumb correlations result in an increased \a1g--\egp--splitting with respect to the LDA, causing a charge transfer that 
pushes \a1g\ spectral weight above the Fermi level.
By computing momentum-resolved spectral functions~\cite{tomczak_vo2_proc,me_vo2,poter_v2o3}, we made explicit predictions for 
angle-resolved photoemission experiments.
Recent measurements  on (V$_{1-x}$Cr$_x$)$_2$O$_3$ ($x$$=$$0.011$)~\cite{arpes_v2o3} nicely agree with the theoretical spectra, further
validating our current understanding of this compound.

\section{Optical properties -- prelude}

An experimental probe complementary to photoemission is optical spectroscopy, 
which is commonly analyzed in terms of the optical conductivity~\cite{millis_review}

\begin{equation}
	\Re\sigma ^{\alpha\beta} (\omega)=\frac{2\pi e ^2\hbar}{V}\sum_{\svek{k}}\int\dif\omega\pr\frac{f({\omega\pr})-f({\omega\pr+\omega})}{\omega} 
\tr\biggl\{ A_{\svek{k}}(\omega\pr+\omega) v_{\svek{k},\alpha} A_{\svek{k}}(\omega\pr)  v_{\svek{k},\beta} \biggr\}
	\label{oc}
\end{equation}
that is given by a convolution of spectral functions $A_{\svek{k}}(\omega)$%
\footnote{In the case of a single orbital, vertex-corrections vanish in the limit of infinite lattice coordination~\cite{PhysRevLett.64.1990}.
Therewith, the response can be expressed in terms of spectral functions.
In the general multi-orbital case this is an approximation.}.
 The transitions are weighted by dipole matrix elements, called Fermi velocities, $v_{\svek{k}, \alpha}$, and the Fermi functions $f({\omega})$ select the range of occupied and empty states, respectively.
$V$ is the unit-cell volume, 
$\alpha$,$\beta$ denote cartesian coordinates, and $\Re\sigma ^{\alpha\beta}$ is the response in $\alpha-$direction for a light polarization along $\beta$.
Both spectral functions and velocities are matrices in the basis
of localized orbitals, which we index by $L=(n,l,m,\gamma)$, with the
usual quantum numbers $(n,l,m)$, while $\gamma$ labels  the individual
atoms in the unit cell.
While the computation of the Fermi velocities is straight-forward \eg in a plane-wave basis, their evaluation becomes tedious
when using localized orbitals as required by many-body approaches such as LDA+DMFT.
To this end, we employ the recently generalized Peierls approach~\cite{me_phd, optic_prl, me_psik}, which we briefly summarize in \ref{app1}.

In their pioneering work, Rozenberg \etal~\cite{PhysRevLett.75.105} analyzed the optical conductivity of V$_2$O$_3$ from the model perspective.
It was concluded that the phenomenology of the temperature dependence
in the conductivity can be understood by appealing to the physics of
the one-band Hubbard model. In the current work, we will substantiate
and extend these observations, based on a realistic multi-band setup. 
 
\section{Optical properties -- results}

Our calculation of the optical conductivity is based on the previous LDA+DMFT electronic structure computation of Ref.~\cite{poter_v2o3}, which used a one-particle Hamiltonian that was downfolded~\cite{Loewdin,nmto} to the vanadium \t2g\ orbitals. Since optical experiments normally probe a much wider frequency range, we employ an upfolding scheme that includes higher energy states on the LDA level. Details are summarized in~\ref{app2}.

\begin{figure}[h!]
\begin{center}
\includegraphics[angle=0,width=0.5\textwidth]{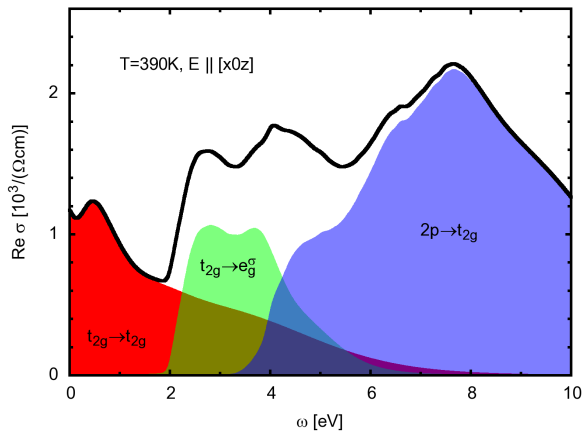}
\caption{Theoretical optical conductivity of V$_2$O$_3$ at $T\approx 390$~K for a light polarization $E\parallel [x\,0\,z]$$=$$[0.13,\, 0.0,\, 0.041]$.
Contributions from different energy sectors (see \ref{app2})~:
\t2g\ $\rightarrow$ \t2g\ (red), \t2g\ $\rightarrow$ \egs\ (green), O$2p$ $\rightarrow$ \t2g\ (blue).  
}
\label{fig1}
\end{center}
\end{figure}

\Fref{fig1} shows our theoretical optical conductivity for 
V$_2$O$_3$ at $T$=390~K for the indicated light polarization.
While in the Kohn-Sham spectrum, the \t2g\ and \egs\ bands are well
separated, the correlations -- accounted for by LDA+DMFT for the \t2g\ 
orbitals only~\cite{poter_v2o3} -- result in an intra-\t2g\ conductivity 
that has weight up to
energies well beyond the onset of transitions into the \egs\ at about 
2~eV. Contributions stemming from transitions from the occupied oxygen 
2p orbitals  into the \t2g\ occur from around 3.5~eV onwards.

\begin{figure}[!ht]
\begin{center}
\includegraphics[angle=-90,width=0.65\textwidth]{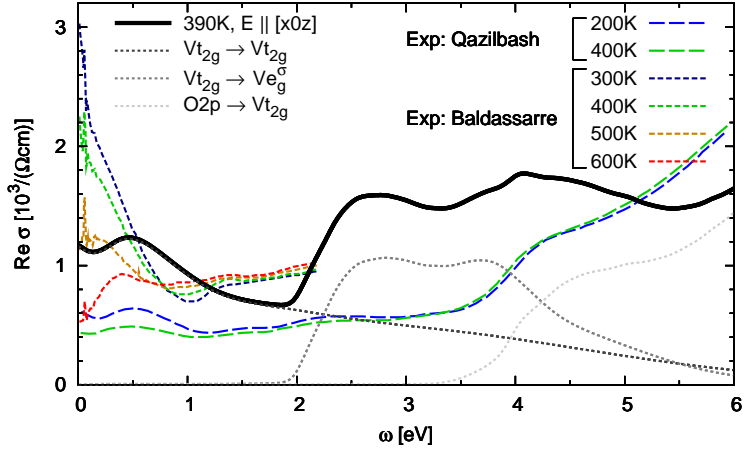}
\caption{Theoretical and experimental conductivity of V$_2$O$_3$. Theory (same polarization as in \fref{fig1})~: total (bold black), orbital contributions (shades of gray, dotted). Experiments~: polycrystalline film, ($T$=$200$, $400$~K) Qazilbash \etal\cite{qazilbash:115121} (blue, green, long dashed), single crystal ($T$=$300$--$600$~K) Baldasarre~\etal\cite{Baldassarre_v2o3} (blue to red, short dashed).
}
\label{fig2}
\end{center}
\end{figure}

Before turning to a more detailed analysis of the different orbital
contributions we compare our results to experimental data 
(see \fref{fig2}).
First, we notice the discrepancies {\it between the experiments}~: 
Recent measurements on single crystals~\cite{Baldassarre_v2o3}
agree well with previous single crystal experiments~\cite{PhysRevLett.75.105}, 
but they are at variance with measurements using a polycrystalline 
film~\cite{qazilbash:115121}. 
While the use of polycrystalline samples especially in a metal might 
be an issue, so is the fact that both single crystal experiments
were performed up to energies of only a few eVs although the extraction 
of the conductivity involves a Kramers-Kronig transform. 
The low energy shape of the theoretical conductivity resembles the 
polycrystalline conductivity, but the absolute values differ.
As to the single crystal one, we note that the order of magnitude compares 
favourably, while the shape tends to be comparable with the high temperature 
curves only.

At high energies, we see that both the onset and the shape of 
oxygen 2p derived contributions agree with experiment. 
The upfolding scheme that uses the 2p bands from LDA is thus a 
good approximation.
This seems not to be true for transitions into \egs\ orbitals~:
Compared with experiment~\cite{qazilbash:115121}, we realize
 that spectral weight is too sharply defined and no identification 
of particular structures is possible. This calls for an LDA+DMFT 
calculation that includes all vanadium 3d orbitals.

We now turn to a detailed analysis of
orbital effects in the optical conductivity. This is a topic that has
not at all been dealt with so far, since previous work neglected inter-band
transitions altogether~\cite{Baldassarre_v2o3}.

At low energy only a small Drude-like tail appears. This can be 
understood from the underlying electronic structure. 
Indeed, the metallic character of V$_2$O$_3$  is mainly a result from 
\a1g\ charge carriers that have spectral weight at the Fermi level 
only in a very limited region of the Brillouin zone (BZ), as can be 
seen in figure 4 of Ref.~\cite{poter_v2o3}. 

As can also be inferred from that work, the local spectral functions 
of \a1g\ and \egp\ character 
display a pseudo-gap-like behaviour, and peak at finite energies 
rather than at the Fermi level, accounting for the feature seen at 0.5~eV 
in the conductivity. 
The latter originates from two types of transitions\footnote{The following 
is inferred from ``momentum resolved optics'', \ie from distinguishing 
contributions of different points in the Brioullin zone (not shown, 
see~\cite{me_phd}).}~:
At energies lower than 0.6~eV the spectral weight is mainly due to
transitions from the \a1g\  into low lying \egp\ orbitals, that are
restricted to a small region in the BZ,  whereas at
slightly higher energies, 0.6~eV and above, contributions are in
majority deriving from \egp\ to \egp\ transitions, which are possible
in a wide region of the BZ, yet are  less prominent
at the $\Gamma$-point. 

At this point we again use our knowledge about the electronic structure
of the compound: Reference \cite{poter_v2o3} 
established an important orbital dependence of the quasi-particle 
coherence scale. 
Indeed, down to 390~K, \egp\ excitations 
are far from being coherent~: The imaginary part of the \egp\ self-energy 
reaches $-0.45$~eV at the Fermi level, while \a1g\ excitations have reached
their coherence regime in our calculation~\cite{poter_v2o3,me_phd}. 
Therewith ($e_g^\pi$) \a1g\ carriers are (not) particularly sensitive 
to changes in temperature.
As discussed above, the low-energy ($<0.6$~eV) optical response
is determined by \a1g--\egp\ transitions, while above 0.6~eV \egp--\egp\
transitions become dominant.
Given these two facts, one can -- even without explicit
calculations -- make some predictions about the behaviour
of the optical response when the temperature is raised:
Upon heating, the purely \egp--derived contributions will not change as much as 
will those that involve the \a1g\ orbitals, so that the low-energy response will be more
sensitive than the weight beyond 0.6~eV. In particular, a broadening 
(and thus reduction in height) is expected for the very low
energy part. 
This gives a natural explanation for the dip behaviour that is 
observed in the experiments when the temperature is raised above
$\sim$ 450~K (see figure 1 in Ref.~\cite{Baldassarre_v2o3} or our \fref{fig2})%
~\footnote{Ref.~\cite{Baldassarre_v2o3} has invoked the change of the
lattice constants upon heating in order to explain this dip.
It is clear that the mechanism based on the orbital-selective
coherence that emerges from our work could not have been 
observed in the calculation of Baldassarre {\it et al}, since, there, inter-band 
transitions were neglected (see their footnote 26 and our discussion in \ref{app1}).
}%
.
Explicit calculations as a function of temperature (including
inter-band transitions) would be desirable to confirm the picture
emerging from our results. This challenging project is left
for future work.

\begin{figure}[!ht]
\begin{center}
\includegraphics[angle=0,width=0.5\textwidth]{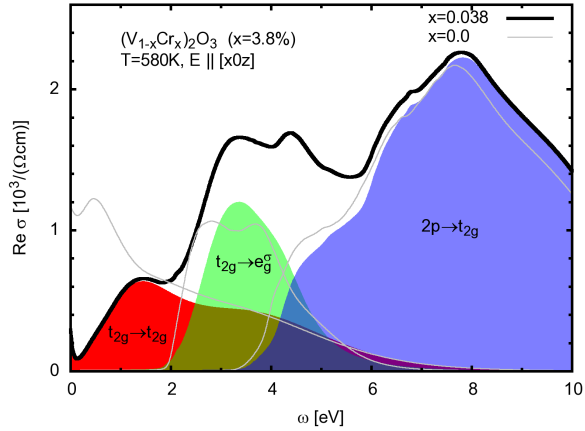}
\caption{Theoretical optical conductivity of $($V$_{1-x}$Cr$_x)_2$O$_3$, x=3.8\% at $T=580$~K for a light polarization $E\parallel[x\,0\,z]$$=$$[0.13, \,0.0,\, 0.0415]$.
Contributions from different energy sectors~:
\t2g\ $\rightarrow$ \t2g\ (red), \t2g\ $\rightarrow$ \egs\ (green), O$2p$ $\rightarrow$ \t2g\ (blue).
Gray lines show the results of pure V$_2$O$_3$ (\fref{fig1}) for comparison.
}
\label{fig3}
\end{center}
\end{figure}

In \fref{fig3} we show theoretical results for the insulator $($V$_{1-x}$Cr$_x)_2$O$_3$ (x=3.8\%).
As discussed previously~\cite{poter_v2o3}, we have low but finite
spectral weight at the Fermi level, which 
results in some optical weight at low frequency.
Unfortunately, no experimental data is available for this composition.
Compared with pure V$_2$O$_3$, we note the suppression of low energy spectral weight and the clear distinction of transitions into the \t2g\ upper Hubbard bands at $\sim$4~eV. 

\section{Conclusions}
In conclusion, we presented calculations of the optical conductivity of V$_2$O$_3$ in its paramagnetic phases using the generalized Peierls approach.
We obtain good agreement with experiment and propose an explanation for the
experimentally witnessed temperature dependence of the response as a
signature
of
the orbital selective coherence of the system.
Our upfolding scheme to include higher energy orbitals captures well the transitions involving oxygen states, but reveal the necessity to 
include all 3d orbitals in an LDA+DMFT electronic structure calculation for this compound.

\ack
We thank M. M. Qazilbash, M. Marsi and A. I. Lichtenstein for discussions on optical and photoemission spectroscopy.
Moreover, we thank our coauthors of Ref.~\cite{poter_v2o3}, 
and and in particular A. I. Poteryaev, for the collaboration  on the electronic structure of
V$_2$O$_3$ that was our starting point.
JMT kindly acknowledges support by the Ecole Polytechnique, where this work
began. 
This work was supported by the French ANR under project
CORRELMAT, and by Idris, Orsay, under project No. 081393.

\appendix
\section{Details of the formalism}
\label{form}

\subsection{Generalized Peierls substitution approach to Fermi velocities}
\label{app1}
The Fermi velocities in \eref{oc} are given by elements of the momentum operator $\mathcal{P}$~:
\begin{eqnarray}
v_{\svek{k},\alpha}^{L\pr L}&=&\frac{1}{m}\bra{\svek{k}L\pr}\mathcal{P}_\alpha\ket{\svek{k}L}\label{a1}
\end{eqnarray}
$L$$=$$(n,l,m,\gamma)$ and $\gamma$ labels atoms in the unit-cell. 
In 
plane waves  \eref{a1} is easily calculated, while using a localized Wannier-like basis 
$\chi_{\svek{R}L}(\vek{r})$=$\braket{\vek{r}}{\vek{R}L}$=$\sum_\svek{k}e^{-\im\svek{k}\svek{R}}\braket{\vek{r}}{\vek{k}L}$
renders the evaluation tedious.
Inspired by the Peierls substitution approach~\cite{millis_review} of lattice models, we can separate the above into~\cite{me_phd,optic_prl,me_psik}~:
\begin{equation}\label{matel2}
v_{\svek{k},\alpha}^{L\pr L}
=\frac{1}{\hbar} \biggl(        \partial_{k_\alpha}\op{H}^{L\pr
                   L}_{\svek{k}} -\im
                   (\rho_{L\pr}^\alpha-\rho_L^\alpha)\op{H}^{L\pr
                   L}_{\svek{k}}        
 \biggr) + \mathcal{F}_{\op{H}}\left[\{\chi_{\svek{R}L}^{\phantom{b}} \}\right]
 \end{equation}
The terms in brackets are the Fermi velocity in the Peierls approximation, which is here generalized to multi-atomic unit-cell~:
$\rho_L^\alpha$ is the $\alpha$--component of the position of atom $\gamma$ within the unit-cell.
This velocity is easy to evaluate since it involves only elements of the Hamiltonian.
Crucial is that $\mathcal{F}_{\op{H}}[\{\chi_{\svek{R}L}^{\phantom{b}} \}]$ reduces to intra-atomic contributions 
in the limit of strongly localized orbitals $\chi_{\svek{R}L}^{\phantom{b}}$~\cite{me_phd}, which makes the generalized Peierls velocity a good approximation for
\eg 3d and 4f systems.

\begin{figure}[!ht]
\begin{center}
\includegraphics[angle=-90,width=0.5\textwidth]{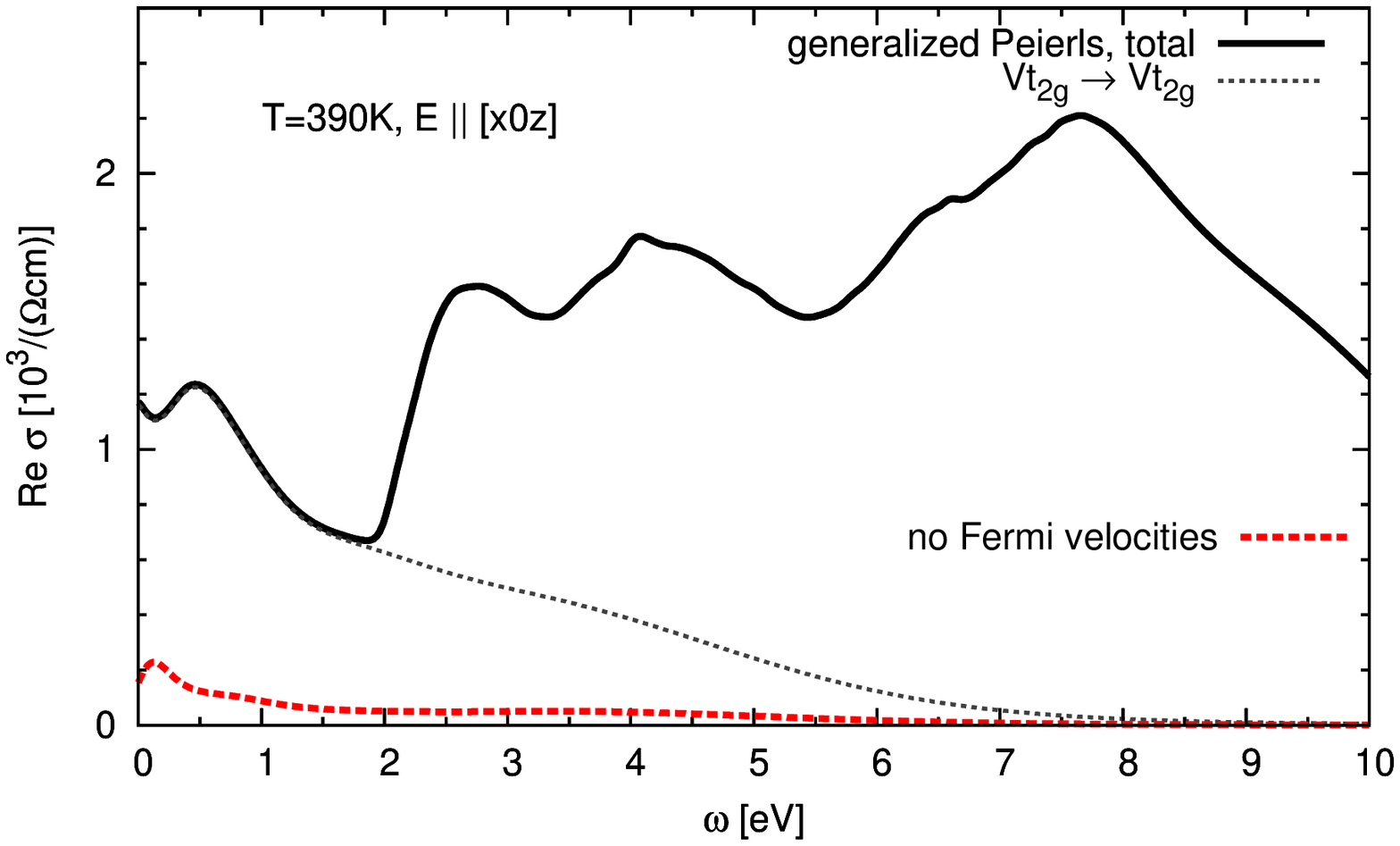}
\caption{Generalized Peierls conductivity versus ``no velocity approximation''.}
\label{fig4}
\end{center}
\end{figure}

Finally, we ask if the computation of the Fermi velocities
is really necessary in practice, or if one could also resort
to a simpler approximation consisting in simply omitting
the Fermi velocities. Due to its simplicity, this approximation 
is in fact relatively popular to obtain qualitative trends of
optical properties in correlated 
systems~\cite{1367-2630-7-1-188,Baldassarre_v2o3}.
Since then the conductivity is then a simple convolution of 
spectral functions, inter-band transitions ($L$$\ne$$L\pr$) are neglected and
intra-band transitions not properly weighted.
As an illustration, we show in \fref{fig4} a comparison of the 
optical conductivity calculated within the generalized Peierls
formalism compared to the one computed from
the simple convolution of spectral functions\footnote{In order to have comparable scales, we chose for the latter case
$v_\svek{k}=r_0 \mathbbm{1}$, with the Bohr-radius $r_0$.}:
Besides the obvious discrepancy in absolute value, omitting the Fermi velocities results in a noticeable change in shape, too.
This is owing to the momentum dependence of the matrix elements that favours certain regions in the Brioullin zone, while attenuating others.

\subsection{Upfolding scheme for higher energy transitions}
\label{app2}
Although 
the treatment of many-body correlations can often be cast into an effective low energy, downfolded
system, the range of its validity is usually far exceeded by optical measurements. Thus it is desirable to allow for optical transitions into higher energy orbitals.  
Also, the computation of the Fermi velocities and the downfolding procedure do not commute~\cite{millis_review}, and hence it makes a difference to which Hamiltonian the generalized Peierls approach is applied. As a matter of fact, the Wannier functions of a full Hamiltonian are more localized than for the downfolded one, whereby the Peierls approximation becomes more accurate\footnote{ 
Indeed the downfolding can be viewed as a unitary transformation that block-diagonalizes the Hamiltonian (see below).
The change in accuracy manifests itself in the basis dependence of the optical conductivity within the Peierls approach.}.
The key quantity for the conductivity is the orbital trace 
in \eref{oc}. For any unitary transformation $U_\svek{k}$ holds
\begin{equation}
\tr\bigl\{ v_\svek{k}A_\svek{k}(\omega\pr)v_\svek{k}A_\svek{k}(\hbox{$\omega\pr$$+$$\omega$})\bigr\}=
\tr\bigl\{U ^\dag_\svek{k} v_\svek{k}U_\svek{k} \widetilde{A}_\svek{k}(\omega\pr)U ^\dag_\svek{k}v_\svek{k}U_\svek{k} \widetilde{A}_\svek{k}(\omega\pr+\omega)\bigr\}\label{trace}
\end{equation}
where we defined $\widetilde{A}_\svek{k}=U ^\dag_\svek{k} A_\svek{k}U_\svek{k}$.
In the case of a pure band-structure calculation (no self-energy), we can choose the transformation such that it performs the downfolding, \ie  
the spectral functions $\widetilde{A}_\svek{k}$  
acquire a block-diagonal form.
We shall distinguish between the low energy ($\mathscr{L}$) and the high energy block ($\mathscr{H}$)~: An LDA+DMFT calculation will
 add local Coulomb interactions only to the former {\it after} the block-diagonalization, which results in a self-energy that lives in this sub-block, while high energy bands 
 remain unchanged and the 
block-diagonality is retained. 
Clearly the downfolding procedure 
is not exact in the many-body framework. 
Indeed the matrices that block-diagonalize the true interacting system also depend on frequency due to the dynamical nature of the self-energy.
Yet, when granting the approximative validity of the downfolding, and using the $U_\svek{k}$ of
the band-structure calculation, we can specify
\begin{eqnarray}
\widetilde{v}_\svek{k}=
\left(
\begin{array}{cc}
V_1 & W\\
W^\dag & V_2\\
\end{array}
\right), 
\widetilde{A}_\svek{k}(\omega\pr)=\left(
\begin{array}{cc}
L & 0\\
0 & H\\\end{array}
\right), 
\widetilde{A}_\svek{k}(\hbox{$\omega\pr$$+$$\omega$})=\left(
\begin{array}{cc}
\bar{L} & 0\\
0 & \bar{H}\\
\end{array}
\right)\label{LH} 
\end{eqnarray}
with $\widetilde{v}_\svek{k}=U ^\dag_\svek{k} v_\svek{k}U^{\phantom{\dag}}_\svek{k}$.
The many-body spectra $L$, $\bar{L}$ are substituted into
the $\mathscr{L}$--sector, while $H$, $\bar{H}$
of the $\mathscr{H}$--sector stem
from the initial band-structure, and
\eref{trace} reads
\begin{equation}\label{dfvkor}
LV_1\bar{L}V_1 + LW\bar{H}W^\dag +
HV_2\bar{H}V_2+HW^\dag\bar{L}W
\end{equation}
For transitions within the $\mathscr{L}$--block, the velocity
 $V_1$ appears, which is the $\mathscr{L}$--block of the {\it transformed} 
  velocity.  
It is different
from the  element computed {\it after} the downfolding. 
With the above, 
we moreover 
include transitions from, to and within the high energy
block\footnote{We can thus distinguish different origins of spectral weight. Yet, we cannot tell apart  contributions within the $\mathscr{L}$--block. While one can suppress selected transitions by setting to zero  Fermi-velocity matrix elements,  contributions
are in that case not additive.}.
Comparing to experiments then  allows to assess whether
correlation effects 
substantially modify also the spectrum of downfolded orbitals, or whether for them the initial band-structure 
is satisfying (see above for the  V$_2$O$_3$ case).

\section*{References}
%\bibliography{../../../../refs,../../../../refs_mine}

\end{document}